\documentstyle[12pt]{article}

   \let\d=\delta \let\e=\epsilon
    \let\k=\kappa
   \let\x=\xi \let\p=\pi 
   \let\f=\phi  
\let\w=\omega \let\G=\Gamma \let\D=\Delta

\let\la=\label  \let\re=\ref
  
\def\nn{\nonumber} \def\bd{\begin{document}} \def\ed{\end{document}}
\def\ds{\documentstyle} \let\fr=\frac \let\bl=\bigl \let\br=\bigr
\let\Br=\Bigr \let\Bl=\Bigl
\let\bm=\bibitem
\let\na=\nabla
\let\pa=\partial \let\ov=\overline
\newcommand{\be}{\begin{equation}}
\newcommand{\ee}{\end{equation}}
\def\ba{\begin{array}}
\def\ea{\end{array}}
\newcommand{\ho}[1]{$\, ^{#1}$}
\newcommand{\hoch}[1]{$\, ^{#1}$}
\newcommand{\bea}{\begin{eqnarray}}
\newcommand{\eea}{\end{eqnarray}}
\newcommand{\ra}{\rightarrow}
\newcommand{\lra}{\longrightarrow}
\newcommand{\Lra}{\Leftrightarrow}
\newcommand{\ap}{\alpha^\prime}
\newcommand{\bp}{\tilde \beta^\prime}
\newcommand{\tr}{{\rm tr} }
\newcommand{\Tr}{{\rm Tr} }
\newcommand{\NP}{Nucl. Phys. }
\newcommand{\tamphys}{\it Center for Theoretical Physics\\
Physics Department \\ Texas A \& M University
\\ College Station, Texas 77843}

\newcommand{\auth}{M. J. Duff\hoch{\dagger} and R. Minasian}

\begin{document}
\begin{titlepage}
\hfill{CTP-TAMU-16/94}

\hfill{hep-th/9406198}

\vspace{74pt}

\begin{center}
{ \large {\bf PUTTING STRING/STRING DUALITY\\
TO THE TEST} }

\vspace{36pt}

\auth

\vspace{10pt}

{\tamphys}

\vspace{54pt}

\underline{ABSTRACT}

\end{center}
After simultaneous compactification of spacetime and worldvolume on $K3$, the
$D=10$ heterotic fivebrane with gauge group $SO(32)$ behaves like a $D=6$
heterotic string with gauge group $SO(28) \times SU(2)$, but with Kac--Moody
levels different from those of the fundamental string. Thus the
string/fivebrane duality conjecture in $D=10$ gets replaced by a string/string
duality conjecture in $D=6$. Since $D=6$ strings are better understood than
$D=10$ fivebranes, this provides a more reliable laboratory in which to test
the conjecture. According to string/string duality, the Green--Schwarz
factorization of the $D=6$ spacetime anomaly polynomial $I_{8}$ into $X_4\,
\tilde{X}_4$ means that just as $X_4$ is the $\sigma$-model anomaly polynomial
of the fundamental string worldsheet so $\tilde{X}_4$ should be the
corresponding
polynomial of the dual string worldsheet. To test this idea we perform a
classical dual string calculation of $\tilde{X}_4$ and find agreement
with the quantum fundamental string result. This also provides an {\it a
posteriori} justification for assumptions made in
 a  previous paper on string/fivebrane duality.
Finally we speculate on the relevance of string/string duality to the
vacuum degeneracy problem.


{\vfill\leftline{}\vfill
\leftline{June 1994}
\vskip 10pt
\footnoterule
{\footnotesize
 \hoch{\dagger} Research supported in part by NSF Grant PHY-9106593
\vskip -12pt}}

\pagebreak
\end{titlepage}
\setcounter{page}{1}

\makeatletter
\@addtoreset{equation}{section}
\makeatother
\renewcommand{\theequation}{\thesection.\arabic{equation}}

\section{Introduction}

In $D$ spacetime dimensions an extended object of worldvolume dimension $d$ is
dual, in the sense of Poincare duality, to another extended object of dimension
$\tilde d$ given by
\be
\tilde d=D-d-2
\la{1}
\ee
the most familiar example being $D=4$, where an electric monopole ($d=1$) is
dual to magnetic monopole ($\tilde d=1$). More recently, attention has focussed
on $D=10$, where a superstring ($d=2$) is dual to a superfivebrane ($\tilde
d=6$) \cite{duff1}--\cite{girardello}.
In this paper, however, we shall focus on $D=6$ where a string ($d=2$) is
dual\footnote{In fact a
string can be dual to another string even in $D<6$ if the fivebrane wraps
around $4$ of the
compactified directions \cite{strominger,khuri2}.}
to another string ($\tilde{d}=2$) \cite{lu5,lu10}.

There is now a good deal of circumstantial evidence to suggest that the string
and the fivebrane theories may actually be different mathematical formulations
of the same physics with the strongly coupled string corresponding to a weakly
coupled fivebrane, and vice-versa. We shall refer to this idea as the
string/fivebrane duality conjecture. In a certain sense, it is a generalization
of the electric/magnetic duality conjecture of \cite{olive}. In previous papers
\cite{lu4,lu5,lu6,dixon2}, it was
pointed out that the Green--Schwarz \cite{green} factorization of the spacetime
anomaly polynomial $I_{12}$ into $X_4 \tilde X_8$ provides a non-trivial check
on this string/fivebrane duality \cite{duff1,strominger}. Just as
$X_4$ is the $\sigma$-model anomaly polynomial of the $d=2$ string worldsheet,
so $\tilde X_8$ should be the corresponding polynomial of the $\tilde d=6$
fivebrane worldvolume. To test this idea, a classical fivebrane calculation of
$\tilde X_8$ was performed \cite{dixon} and found to be in agreement with the
string one-loop result, thus supporting the conjecture.  Although the fivebrane
calculation of the {\it Yang--Mills} contribution to $X_8$ is relatively
straightforward, however,
an educated guess had to be made for the {\it gravitational} contributions; a
guess which has
recently been questioned in \cite{izquierdo,blum}.

In this paper we repeat the test for string/string duality in
$D=6$. Now the factorization of the spacetime anomaly polynomial $I_{8}$
into $X_4 \tilde X_4$ means that $\tilde X_4$ should be the
$\sigma$-model anomaly polynomial of
the $\tilde d=2$ dual string worldsheet. Since $D=6$ strings are
better understood than $D=10$ fivebranes, this provides a more
reliable theoretical laboratory in which to test the conjecture.
In particular, the gravitational contributions to the $\sigma$-model
anomaly for a string are well-known, thus avoiding the guesswork \cite{dixon}
that had to be made for those of the fivebrane.
We perform a classical dual string calculation of $\tilde X_4$
and once again find agreement with the fundamental string
one-loop result\footnote{Since the result of \cite{dixon} is
sometimes misunderstood, we stress that two {\it independent} calculations of
$\tilde X_8$ were
being compared. The first was the fundamental string one-loop calculation of
the spacetime anomaly
\cite{green}.  The second was the fundamental fivebrane tree-level calculation
of the worldvolume
anomaly \cite{dixon}.  This latter result is obtained from the fundamental
superfivebrane
$\sigma$-model action of \cite{bergshoeff2} augmented by the internal $SO(32)$
degrees of freedom
\cite{dixon1,dixon2}. Being tree-level, such a calculation is possible in spite
of our ignorance
of how to quantize the fivebrane.  This is, of course, quite different from
calculating the string
spacetime polynomial \cite{salam,gates} from the dual supergravity theory
\cite{chamseddine,lu4}
obtained by dualizing the usual string effective action
\cite{bergshoeff,chapline,candelas}, which is
guaranteed by construction to yield the same $\tilde X_8$.  Similarly, in the
present paper, we will
compare two {\it independent} calculations of $\tilde X_4$.}.

The chiral, anomaly-free, fundamental string we shall consider will be the
heterotic string obtained by spacetime compactification on $K3$ of the $SO(32)$
heterotic string in $D=10$. The resulting $D=6$ string has gauge group $SO(28)
\times SU(2)$ with Kac--Moody levels $k_{SO(28)}=1$ and $k_{SU(2)}=1$. The dual
string is obtained from the fivebrane by simultaneous compactification of
spacetime
and worldvolume on $K3$. As discussed in section \ref{K3}, this string is
also chiral and has the same gauge group but with different Kac--Moody levels:
$\tilde k_{SO(28)}=4$ and $\tilde k_{SU(2)}=-20$ (the minus sign corresponds to
the $SU(2)$
worldsheet fermions having the opposite chirality). The calculation of $\tilde
X_4$ can be done either directly with the $D=6$ dual string or else by starting
with the $D=10$ fivebrane and using
\be
\tilde X_4=\int_{K3} \tilde X_8
\la{1a}
\ee
Both ways yield the same result, and this provides an {\it a posteriori}
justification for the
gravitational contributions to $\tilde X_8$ assumed in \cite{dixon}.

\section{Review of string/string duality}

We begin by recalling some facts about duality in $D=6$ \cite{lu10}. There
are two formulations of $D=6$ supergravity, both with a 3-form field
strength\footnote{There is
also a {\it self-dual} supergravity in D=6 and an associated self-dual string
\cite{lu10} for
which $H_3=*H_3$.  While interesting in its own right, this is not the subject
of the present
paper.} The bosonic part of the usual action takes the form
\be
I_6=\frac{1}{2\kappa^2}\int d^6x
\sqrt{-g}e^{-2\Phi}[R+4(\partial\Phi)^2-\frac{1}{2.3!}H_3{}^2+...]
\la{a}
\ee
To within Chern--Simons corrections to be discussed below, $H_3$ is the curl of
a 2-form $B_2$
\be
H_3=dB_2+...
\la{b}
\ee
The metric $g_{\mu\nu} (\mu=0,...,5)$ is related to the canonical Einstein
metric $g^c{}_{\mu\nu}$ by
\be
g_{\mu\nu}=e^{\Phi}g^c{}_{\mu\nu}
\la{metric}
\ee
where $\Phi$ the $D=6$ dilaton.\footnote{Our use of the symbol $\Phi$ conforms
with the notation of section 6 of \cite{lu10} where $8\Phi= (D - 2)\alpha
\phi$ and where $\alpha =\sqrt{2}$ for $D=6$.} Similarly, the dual supergravity
action is given by
\be
\tilde I_6=\frac{1}{2\kappa^2}\int d^6x \sqrt {-\tilde g}e^{2\Phi}[\tilde
R+4(\partial\Phi)^2-\frac{1}{2.3!}\tilde H^{}_3{}^2+...]
\la{c}
\ee
To within Chern--Simons corrections to be discussed below, $\tilde H_3$ is
also the curl of a 2-form $\tilde B_2$ %
\be
\tilde H_3=d\tilde B_2+...
\la{d}
\ee
The metric $\tilde g_{MN}$ is related to the canonical Einstein metric by
\be
\tilde g_{\mu\nu}=e^{-\Phi}g^c{}_{\mu\nu}
\la{e}
\ee
The two supergravities are related by Poincare duality:
\be
\tilde H_3 = e^{-2\Phi}\, {\ast H_3}
\la{3}
\ee
where $\ast$ denoted the Hodge dual. (Since this equation is conformally
invariant, it is not necessary to specify which metric is chosen in forming the
dual.) This ensures that the roles of field equations and Bianchi identities in
the one version of supergravity are interchanged in the other. As field
theories, each supergravity seems equally as good. In particular,
provided we couple them to $SO(28) \times SU(2)$ super Yang-Mills, then both
are anomaly-free.\footnote{There are other many other anomaly-free groups in
$D=6$ \cite{salam,randj} but this is the one we shall focus on in this paper.}
Since the usual version corresponds to the field theory limit of the heterotic
string, it was natural to conjecture \cite{lu5,lu10} that other version
corresponds to the field theory limit of a ``dual string''.

The bosonic action of the fundamental $D=6$ heterotic string
is given by
\be
 S_2 = {1 \over 2\p \ap} \int_{M_2} d^2 \x \left(
-\frac{1}{2}\sqrt{-\gamma}\gamma^{ij}\partial_iX^{\mu}\partial_jX^{\nu}g_{\mu\nu}
-{1\over 2}
\e^{ij} \partial_iX^{\mu}\partial_jX^{\nu} B_{\mu\nu} +...\right)
\la{1b}
\ee
where $\x ^i$ ($i=1,2$) are the worldsheet coordinates,
$\gamma_{ij}$ is the worldsheet metric and $(2\p\ap)^{-1}$ is the string
tension $T$. The metric and 2-form appearing in $S_2$ are the same as those in
$I_6$. This means that under the rescalings with constant parameter $\lambda$:
\bea
g_{\mu\nu}& \rightarrow &\lambda^2 g_{\mu\nu} \nn \\
B_{\mu\nu}& \rightarrow &\lambda^2 B_{\mu\nu} \nn \\
e^{2\Phi}& \rightarrow &\lambda^2 e^{2\Phi} \nn \\
\gamma_{ij}& \rightarrow &\lambda^2 \gamma_{ij}
\la{1c}
\eea
both actions scale in the same way
\bea
I_6& \rightarrow &\lambda^2 I_6 \nn \\
S_2& \rightarrow &\lambda^2 S_2
\la{1d}
\eea

We therefore assume that the bosonic action of the dual $D=6$ heterotic
string given by
\be
\tilde S_2 = {1 \over 2\p \tilde\ap} \int_{\tilde M_2} d^2 \tilde\x \left(
-\frac{1}{2}\sqrt{-\tilde\gamma}
\tilde\gamma^{ij}\partial_iX^{\mu}\partial_jX^{\nu}\tilde g_{\mu\nu}
-{1\over 2} \e^{ij} \partial_iX^{\mu}\partial_jX^{\nu}\tilde B_{\mu\nu}
+...\right)
\la{1e}
\ee
where $\tilde \x^i$ ($i=1,2$) are the dual
worldsheet coordinates, $\tilde \gamma_{ij}$ is the dual worldsheet
metric and $(2\p\tilde \ap)^{-1}$ is the dual string tension $\tilde T$.
Furthermore we assume that the metric and 2-form appearing in $\tilde S_2$ are
the same as those appearing in $\tilde I_6$. The virtue of these
identifications is that under the recalings with constant parameter $\tilde
\lambda$:
\bea
\tilde g_{\mu\nu}&\rightarrow &\tilde \lambda^2\tilde g_{\mu\nu} \nn \\
\tilde B_{\mu\nu}&\rightarrow &\tilde \lambda^2 \tilde B_{\mu\nu} \nn \\
e^{-2\Phi}&\rightarrow &\tilde \lambda^{2}e^{-2\Phi} \nn \\
\tilde \gamma_{ij}& \rightarrow &\tilde \lambda^2 \tilde \gamma_{ij}
\la{1f}
\eea
both actions again scale in the same way:
\bea
\tilde I_6&\rightarrow &\tilde \lambda^2 \tilde I_6 \nn \\
\tilde S_2&\rightarrow &\tilde \lambda^2\tilde S_2
\la{1g}
\eea
The duality relation (\ref{3}) is invariant under both rescalings.  Since the
dilaton enters the dual string equations with the opposite sign to the
fundamental string, it was argued in \cite{lu5,lu10} that the strong coupling
regime of the string should correspond to the weak coupling regime of the dual
string:
\be
{\rm g}_{\rm dual~string} = <e^{-\Phi}>={\rm g}_{\rm string}^{-1}
\ee
\la{6}
where ${\rm g}_{\rm string}$ and ${\rm g}_{\rm dual~string}$ are the string
and dual string loop expansion parameters.

Further evidence for this duality is
provided by the complementary discovery that the dual string emerges as
a soliton solution of the fundamental string, and vice-versa
\cite{lu10}. The combined supergravity-source action $I_6+S_2$ admits the
singular
elementary string solution \cite{dabholkar}
\bea
ds^2&= &(1-a^2/r^2)(-d\tau^2+d\sigma^2) + (1-a^2/r^2)^{-1}dr^2 +(1-a^2/r^2)
r^2d\Omega_{3}{}^2\\
e^{2\Phi}&=&1-a^2/r^2\\
e^{-2\Phi}*H_3&=&2a^2\epsilon_3
\eea
\la{elem}
where
\be
a^2=\kappa^2 T/\Omega_3
\ee
and $\Omega_3$ is the volume of $S^3$.  The source-free action $I_6$ also
admits the non-singular
solitonic string solution \cite{lu10}
\bea
ds^2&= &-d\tau^2+d\sigma^2 + (1-\tilde a^2/r^2)^{-2}dr^2 + r^2d\Omega_{3}{}^2\\
e^{-2\Phi}&=&1-\tilde a^2/r^2\\
H_3&=&2\tilde a^2\epsilon_3
\eea
\la{soliton}
whose tension $\tilde T$ is given by
\be
\tilde a^2=\kappa^2 \tilde T/\Omega_3
\ee
The Dirac quantization rule relating the Noether ``electric'' charge
\be
e=\frac{1}{\sqrt{2}\kappa}\int_{S^3}e^{-2\Phi}*H_3
\ee
to the topological ``magnetic'' charge
\be
g=\frac{1}{\sqrt{2}\kappa}\int_{S^3}H_3
\ee
translates into a quantization condition on the two tensions \cite{lu2}:
\be
2 \k^2 = n (2\p)^3 \ap \tilde \ap, \qquad \, n={\rm integer}
\la{6a}
\ee
Similarly, the dual supergravity-source action $\tilde I_6 +\tilde S_2$ admits
the dual string as
the fundamental solution and the fundamental string as the dual solution.  When
expressed in terms
of the dual metric (\ref{e}), however, the former is singular and the latter
non-singular.
Both the string and dual string soliton solutions break half
the supersymmetries, both saturate a Bogomol'nyi bound between the mass and the
charge. These solutions are the
extreme mass equals charge limit of more general two-parameter black string
solutions \cite{horowitz,lu10}.

Duality mixes up string and dual string loops in the sense that
the roles of worldsheet loop expansions and spacetime loop expansions are
interchanged \cite{lu5,lu10}. Consequently what is a quantum effect for the
string might be a classical effect for the dual string, and vice versa. At
higher loop orders, this leads to an infinite number of
non-renormalization theorems (including the vanishing of the cosmological
term) all of which are consistent with known string calculations to higher
orders both in $\ap$ (worldsheet loops) and ${\rm g}_{\rm string}$ (spacetime
loops). It is this loop mixing which allows us to test
string/string duality, in spite of our ignorance of how to quantize the dual
string. If duality is correct, we should be able to reproduce string loop
effects from tree-level dual strings! In this paper we shall show in
particular how to reproduce the Green--Schwarz spacetime anomaly corrections to
the $H_3$ field equations (a fundamental string one-loop effect) from the
Chern--Simons worldsheet anomaly corrections to the $\tilde H_3$ Bianchi
identities (a dual string tree-level effect).

\section{The fundamental string}
\label{K3}

The fundamental string we shall consider will be the $D=6$ heterotic string
obtained by
compactification of the $D=10$ heterotic string on $K3$. We choose this because
it is essential
for our purposes that both the string and the dual string be chiral.

$K3$ is a four-dimensional compact closed simply-connected manifold. It is
equipped with a self-dual metric and hence its holonomy group is $SU(2)$. It
entered the physics
literature as a gravitational instanton \cite{hawking,page},
its Pontryagin number being %
\be
p_1=-\frac{1}{8\pi^2}\int_{K3}trR^2=-48
\la{23}
\ee
However, it was then invoked in a Kaluza-Klein context in
\cite{duff6,townsend}
where it was used, in particular, as a way of compactifying $D=10$ supergravity
to $D=6$. Half the spacetime supersymmetry remains unbroken as a consequence of
the $SU(2)$ holonomy, and hence it gives rise to an $N=1$ supergravity in
$D=6$.

There are four $N=1,D=6$ supermultiplets to consider:
$$
\begin{array}{ll}
Supergravity~~~~~~&g_{\mu\nu}, \Psi^A{}_L{}_{\mu}, B_L{}_{\mu\nu}\\
Tensor~~~~~~~~~~~~&B_R{}_{\mu\nu}, \chi^A{}_R, \phi\\
Hypermatter~~~~~~~&\psi^a{}_R, \phi^{\alpha}\\
Yang-Mills~~~~~~~~&A_{\mu}, \lambda^A{}_L
\end{array}
 $$
The scalars $\phi^{\alpha}$ parametrize a quaternionic Kahler manifold of the
form ${\cal G}/{\cal H} \times Sp(1)$. The index $A=1,2$ labels $Sp(1)$ and
the index $a$ lables one of the representations of $\cal H$. All spinors are
symplectic Majorana--Weyl. The 2-forms $B_L{}_{\mu\nu}$ and $B_R{}_{\mu\nu}$
have 3-form field strengths that are self-dual and anti-self-dual,
respectively. Only with the combination of one supergravity multiplet and one
tensor multiplet do we have a conventional covariant Lagrangian
formulation. In the case of $K3$, the massless sector of the $D=6$ theory
coming from the
supergravity multiplet in $D=10$ consists of this combination plus $20$
hypermatter multiplets. The $80$ scalars belong to the coset $SO(20,4)/{SO(20)
\times SO(3) \times Sp(1)}$ \cite{duff8,seiberg,aspinwall}, this being the
moduli space of $K3$. There are no vector multiplets since $K3$ has no
isometries and is simply connected.

This argument was generalized to supergravity-Yang-Mills with gauge groups
$G_{10}=SO(32)$ or $G_{10}=E_8\times E_8$ in \cite{green2}, where it was
emphasized that the resulting $N=1,D=6$ supergravity would be chiral and
anomaly free\footnote{$K3$ was thus the forerunner of Calabi-Yau
compactification \cite{candelas} from $D=10$ to $D=4$. $K3$ has a
curious way of cropping up in the superstring literature in a variety of
apparently unrelated contexts
\cite{duff6,townsend,seiberg,aspinwall,green2,gellmann,duff7,erler,gauntlett}.}.
Counting the number of $D=6$ multiplets coming from the Yang-Mills sector is
more subtle. As discussed in \cite{witten}, the values of the background
fields, $R_0$ and $F_0$, are not independent. The requirement that $H_3$ be
globally defined leads, on integrating (\ref{11x}), to the constraint
\be
\int_{M_4}(tr_{SO(32)}F_0{}^2-tr_{SO(1,9)}R_0{}^2) = 0
\la{24}
\ee
for any closed sub-manifold $M_4$ of the ten-dimensional spacetime. In this
background the effective lower-dimensional theory has a reduced gauge symmetry.
If the non-zero fields $F_0$ span a subgroup $H \subset G_{10}$ then the gauge
group in the lower dimension will be given by $G$ such that
\be
G_{10} \supset G \times H
\la{25}
\ee
The adjoint representation of $G_{10}$ can be decomposed into a sum of
representations
\be
adj~G_{10}= \sum_{i} (L_i, C_i)
\la{26}
\ee
where $L_i$ and $C_i$ are irreducible representations of $G$ and $H$
respectively. In particular, for $G_{10}=SO(32)$,
\be
\sum_{i}dim~L_i \cdot dim~C_i=dim~G_{10}=496
\la{27}
\ee
In the present context we choose $M_4$ to be $K3$ and the backgrounds $R_0$
and $F_0$ to lie only in the four compactified directions. Specifically we
``embed the the holonomy group in the gauge group'' \cite{charap,candelas} by
taking $F_0$=$R_0$ to lie in the $SU(2)$ subgroup\footnote{Different embeddings
were chosen in \cite{green2}.} of $SO(32)$. Then the Yang-Mills supermultiplets
are those of $G=SO(28) \times SU(2)$. Thus

$$
\begin{array}{ll}
L_1=(378,1)&C_1=1\\
L_2=(1,3)&C_3=1\\
L_3=(28,2)&C_2=2\\
L_4=(1,1)&C_4=3
\la{28}
\end{array}
$$
The number of left-handed spinor mutiplets in the representation $L_i$ of $G$
is given by an index theorem:
\be
N_i{}^{1/2}=\frac{1}{8\pi^2}\int_{M_4}[-\frac{1}{2}tr_{C_i}F_0{}^2
+\frac{1}{48}dim~C_i{}trR_0{}^2]
\la{29}
\ee
In the case of $K3$, on using (\ref{24}) and (\ref{25}), this reduces to
\be
N_i{}^{1/2}=(dim~C_i-12R_i)
\la{30}
\ee
where $R_i$ is the ratio
\be
R_i=\frac{tr_{C_i}F_0{}^2}{trF_0{}^2}
\la{31}
\ee
So coming from the Yang-Mills sector in $D=10$ we have
\bea
N_{(378,1)}{}^{1/2}&=&1-0=1 \nn \\
N_{(1,3)}{}^{1/2}&=&1-0=1 \nn \\
N_{(28,2)}{}^{1/2}&=&2-12=-10 \nn \\
N_{(1,1)}{}^{1/2}&=&3-48=-45,
\la{32}
\eea
the first two being the adjoint representation left-handed gaugino
superpartners of the $G=SO(28) \times SU(2)$ gauge fields belonging to the
Yang-Mills supermultiplets, and the last two being the right-handed
superpartners of $\phi^a$ in the hypermatter multiplets.

\section{Worldsheet and spacetime anomalies}

Let us first consider the $G$ Yang--Mills and Lorentz
Chern--Simons corrections to $H_3$. Let us define $F = dA + A^2$
where the gauge fields
$A = A_M\, dx^M$ are matrices in the adjoint representation
of the gauge groups and $R = d\omega + \omega^2$ where the Lorentz connections
$\omega = \omega_M\,dx^M$ are in the vector representation.
Let us further define \cite{warner}
\bea
I_4 &=& \frac{1}{2(2\p)^2} \Bl [ -\sum_l \frac{k_l}{g_l}\Tr F_l^2
+ \tr R^2 \Br ] \nonumber \\
d \w_{3} &=& I_{4} \nonumber \\
 \d \w_{3} &=& d \w^1_{2}
\la{9}
\eea
where the sum is taken over the gauge groups appearing in
$G= G_1 \times G_2 \times \cdots \times G_l$,
$k_l$ are the corresponding levels of the Kac--Moody algebra and $g_l$ are the
dual Coxeter numbers.
Then the action $S_2$ can be modified so as to be both gauge invariant and
Lorentz invariant
provided \cite{hull}
\be
\d B_2 = \frac{1}{2} \ap (2\p)^2 \w^1_2
\la{10}
\ee
and hence the gauge invariant field strength is given by
\bea
H_3 &=& dB_2 - \frac{1}{2} \ap (2\p)^2 \w_3 \nonumber \\
dH_3 &=& - \frac{1}{2} \ap (2\p)^2 I_4
\la{11}
\eea
In the case of $K3$
compactification of the $SO(32)$ string, we have
\be
SO(32) \supset SO(28) \times SU(2)
\la{38}
\ee
and the fundamental representation decomposes as
\be
32 \rightarrow (28,1) + (1,2) + (1,2)
\la{39}
\ee
so the worldsheet fermions carrying the $SO(28)\times SU(2)$ symmetry
correspond to Kac--Moody levels $k_{SO(28)}=1$ and $k_{SU(2)}=1$. Similarly,
the gravitational anomaly is as in $D=10$ but where $R$ now belongs to
$SO(1,5)$ instead of $SO(1,9)$. Thus the Yang-Mills and Lorentz Chern--Simons
corrections are given by
\bea
dH_3=-\frac{\ap}{4}\Br[ -\tr F_{SO(28)}{}^2 - 2\tr F_{SU(2)}{}^2
+ \tr R^2 \Br ]
\la{9a}
\eea
where we have used $TrF_{SO(28)}{}^2=26trF_{SO(28)}{}^2$,
$TrF_{SU(2)}{}^2=4trF_{SU(2)}{}^2$ and
$g_{SO(28)}=26$, $g_{SU(2)}=2$. This modification to the Bianchi identity is
thus a classical string effect (i.e. tree level in the $D=6$ string loop
expansion). This is
confirmed by the observation that there is no dilaton dependence and that
$dH_3$ is independent of
$\k^2$ and that the $2\pi$ factors cancel out.  By supersymmetry, the same
combination of Yang-Mills
field strengths and curvatures appearing in (\ref{9a}) also appears in the
$D=6$ tree level
action
\bea
S^{(0)}{}_{YM}=\frac{1}{2\kappa^2}\int d^6
x\sqrt{-g}e^{-2\Phi}\frac{\ap}{8}t^{\mu\nu\rho\sigma} \Br[
-\tr F_{SO(28)}{}_{\mu\nu}F_{SO(28)}{}_{\rho\sigma}\nn\\
 - 2\tr
F_{SU(2)}{}_{\mu\nu}F_{SU(2)}{}_{\rho\sigma}
 + \tr R_{\mu\nu} R_{\rho\sigma} \Br ]
\la{9c}
\eea
where
\bea
t^{\mu\nu\rho\sigma}=\frac{1}{2}(g^{\mu\rho}g^{\nu\sigma}-g^{\mu\sigma}g^{\nu\rho})
\la{t}
\eea

Next we turn to the Green--Schwarz anomaly cancellation mechanism \cite{green}.
The spacetime anomaly polynomial $I_8$ of this $D=6$ string has been
calculated by Erler \cite{erler} who finds, as expected, that it factorizes in
the form %
\be
I_8=X_4\tilde X_4
\la{33}
\ee
where
\be
X_4=\frac{1}{2}I_4 =\frac{1}{4(2\p)^2} \Bl [ -\tr F_{SO(28)}{}^2 -2\tr
F_{SU(2)}{}^2
+\tr R^2 \Br ]
\la{34}
\ee
and
\be
\tilde X_4=\frac{1}{4(2\p)^2} \Bl [ -2\tr F_{SO(28)}{}^2 +44\tr
F_{SU(2)}{}^2 - \tr R^2 \Br ]
\la{35}
\ee
As a consistency check, one notes that factorization requires the absence of a
$trR^4$ term in $I_8$ and that this is guaranteed if \cite{green2}
\be
\sum_{i=1}^{4} N_i{}^{1/2}dim~L_i=-224
\la{37}
\ee
as may be verified from (\ref{32}).

 Defining $\tilde \omega_3$ by $\tilde X_4=\frac{1}{2} d \tilde \omega_3$, the
consistent anomaly is then cancelled by adding to the effective action
\be
\D \G_2 = - 2\p \int_{M_{6}} \Bl ( \frac{1}{ \ap (2\p)^2} \, B_2\,
 \tilde X_4 + \frac{1}{3}\, \w_3 \,\tilde \w_3 \Br )
\la{15}
\ee
and recalling the transformation rule for $B_2$ given in (\re{10}). Now
with the normalization of the kinetic term for $B_2$ given in (\ref{a}),
the addition of (\ref{15}) modifies the field equation to
\bea
d(e^{-2\Phi}\, \ast H_3) &=& \frac{2\k^2}{\ap (2\p)}\, \tilde X_4 \nn \\
&=& \frac{\k^2}{2(2\p)^{3} \ap}\, \Bl [ -2\tr F_{SO(28)}{}^2 +44\tr
F_{SU(2)}{}^2 - \tr R^2 \Br ]
\la{17}
\eea
This modification to the field equations is thus a string one-loop effect.
This is confirmed by the dilaton dependence on the left hand side and by noting
that the right hand side is linear in $\k^2$ and involves a factor
$1/({2\pi})^3$ appropriate to a one-loop Feynman integral in $D=6$. By
supersymmetry, the same
combination of Yang-Mills field strengths and gravitational curvatures
appearing in
({\ref{17}) also appears in the $D=6$ one loop action
\bea
S^{(1)}{}_{YM}=\int d^6 x\sqrt{-g}\frac{1}{8(2\pi)^3\ap}t^{\mu\nu\rho\sigma}
\Br[
-2\tr F_{SO(28)}{}_{\mu\nu}F_{SO(28)}{}_{\rho\sigma}\nn\\
 +44\tr
F_{SU(2)}{}_{\mu\nu}F_{SU(2)}{}_{\rho\sigma}
 - \tr R_{\mu\nu} R_{\rho\sigma} +...\Br ]
\la{17b}
\eea

After this summary of the $D=6$ string, we are now in a position to test
string/string duality by reproducing (\ref{17}) and (\ref{17b}) from the
classical dual string.

\section{String/string test in $D=6$}

After compactification to $D=6$, a $D=10$ fivebrane will appear as a fivebrane,
a fourbrane, a threebrane, a membrane or a string according as it wraps around
$0,1,2,3$ or $4$ of the compactified directions.  The fivebrane is trivial
having no degrees of
freedom in $D=6$.  The dual supergravity theory in $D=6$  obtained by
compactifying the dual $D=10$
supergravity on $K3$, will consist of the same combination of supergravity
multiplet and one tensor
multiplet as in the fundamental theory but instead of $20$ hypermultiplets,
there  will be $19$
linear multiplets of type $1$ and one of type $2$.  In a type $1$ linear
multiplet \cite{bergshoeff3}
one of the four scalars $\phi$ (an $SU(2)$ singlet) is swapped for a $4$-form
$b_{\mu\nu\rho\sigma}$;
in a type $2$ linear multiplet three of the four scalars (an $SU(2)$ triplet)
are swapped for three
$4$-forms. The appearance of these $4$-forms means that the dual string alone
cannot be responsible
for this low-energy limit; the threebrane is also contributing.  Only the
odd-branes will
display $sigma$-model anomalies \cite{dixon1,dixon} and we might therefore
expect contributions to the $D=6$ spacetime anomaly polynomial $I_8$ of the
form $X_0 \tilde X_8$, $X_2 \tilde X_6$, and $X_4 \tilde X_4$. However, $X_0$
is trivially zero and $X_2$ will involve $trR$ which is zero and $trF$ which is
also zero since our gauge group has no abelian factors. For the purposes of
the $sigma$-model anomaly, therefore, we may focus just on the the massless
states of the dual string worldsheet. These follow from the compactification of
the fivebrane worldvolume $\tilde M_6$ to $\tilde M_2 \times K3$. The
$\kappa$-symmetric,
spacetime supersymmetric fivebrane action (in the absence of internal symmetry)
has been constructed in \cite{bergshoeff2} using the Green--Schwarz variables
$x^\mu,\Theta^\alpha$ with $\mu=0,...,9$ and $\alpha=1,...,16$. Since the
scalar d'alembertian just splits as $\Delta_{6}= \Delta_{2} + \Delta_{K3}$, the
number of massless scalar fields is unchanged under compactification. So the
Green--Schwarz variables $x^\mu$ will remain the same, except that only six of
them will be regarded as spacetime coordinates. Similarly, by spacetime
supersymmetry, the $\Theta^\alpha$ which transformed as a $16$ of $SO(1,9)$ are
now interpreted as a $(4,4)$ of $SO(1,5) \times SO(4)$. This counting is
exactly the same as that of the fundamental Green--Schwarz string. Hence, the
contribution to the gravitational sigma-model anomaly will be the same as that
of the fundamental string.

A complete covariant $\kappa$-symmetric Green--Schwarz action for the {\it
heterotic} fivebrane (i.e. with the internal symmetry included) is still
lacking. If the field theory limit of the heterotic fivebrane is indeed to
coincide with the $SO(32)$ anomaly-free supergravity, however, we must include
this internal symmetry in some way. In \cite{dixon}, we suggested two ways that
would yield the same $\sigma$-model anomaly and that mimic the heterotic
string: Weyl fermions in the fundamental of $SO(32)$ or a level $1$ WZW model.
In this paper, we focus on the fermionic formulation; in a subsequent paper we
will
arrive at the same conclusions by deriving the dual string Kac-Moody algebra
from the fivebrane Mickelsson-Faddeev algebra \cite{minasian}.

The calculation of the spin $1/2$ fermions on the dual string
worldsheet parallels that of section 3. The fundamantal representation of
$G_{10}$ can be decomposed into a sum of representations
\be
fund~G_{10}= \sum_{i} (l_i, c_i)
\la{26ws}
\ee
where $l_i$ and $c_i$ are irreducible representations of
$G=SO(28) \times SU(2)$ and $H=SU(2)$ respectively. The number of left-hand
spinor multiplets in the representation $l_i$ of $G$ is given by an index
theorem:
\bea
n_i{}^{1/2} &=& \frac{1}{4\pi^2}\int_{M_4}[-\frac{1}{2}tr_{c_i}F_0{}^2
+\frac{1}{48}dim~c_i~trR_0{}^2] \nn \\
&=& 4(dim~c_i-12~r_i)
\la{29ws}
\eea
where $r_i$ is the ratio
\be
r_i=\frac{tr_{c_i}F_0{}^2}{trF_0{}^2}
\ee
This differs from (\ref{29}) by a factor of 4 since we are going from Weyl
fermions in $\tilde d = 6$ to Majorana--Weyl in $\tilde d = 2$, as opposed to
Majorana--Weyl in $D = 10$ to symplectic Majorana--Weyl in $D = 6$.
Since, under $SO(32) \rightarrow SO(28) \times SU(2) \times SU(2)$,
\be
32 \rightarrow (28,1,1) + (1,2,2)
\la{39ws}
\ee
we have
\bea
n_{(28,1)}{}^{1/2}&=&4(1-0) = 4 = \tilde k_{SO(28)} \nn \\
n_{(1,2)}{}^{1/2}&=&4(2-12) = -40 = 2 \tilde k_{SU(2)}
\la{32ws}
\eea
where $\tilde k_{SO(28)}=4$ and $\tilde k_{SU(2)}=-20$ are the dual Kac--Moody
levels\footnote{As far as we are aware, the numbers (4,20) appearing here have
no direct connection to the numbers (4,20) describing the moduli space of
$K3$.}.  So in analogy with (\ref{11}) the complete sigma-model anomaly may be
written\footnote{The change of sign relative to (\ref{11}) is chosen so as to
accord with the fivebrane conventions of \cite{dixon} and Appendices.}
\be
d \tilde H_3 = \frac{1}{2} \tilde \ap (2\p)^2 \tilde I_4
\la{11y}
\ee
where from (\ref{9})
\be
\tilde I_4=\frac{1}{2(2\p)^2} \Bl [ -4\tr F_{SO(28)}{}^2 +40\tr
F_{SU(2)}{}^2 + \tr R^2 \Br ]
\la{35yy}
\ee
(Note that since $\tilde I_4$ is independent of the metric, there is no need
for tildes on the
field strengths or curvatures.)  The negative sign in the $SU(2)$ case
corresponds to the opposite
chirality of fermions on the worldsheet.  This is not yet in agreement with $2
\tilde X_4$ of (\ref
{35}) but previous experience with the fivebrane in $D = 10$ suggests that we
should not expect
it to be \cite{dixon}. First we note that
\be
\frac{1}{2} \tilde I_4 - \tilde X_4 = \frac{1}{2(2\p)^2}
\Bl [ -\tr F_{SO(28)}{}^2 - 2\tr F_{SU(2)}{}^2 + \tr R^2 \Br ] = 2 X_4
\la{35xy}
\ee
so that the discrepancy is twice the sigma-model anomaly of the fundamental
string. Up until now we have been assuming that the supergravity $\tilde B_2$
appearing in (\ref{d}) should be identified with the dual string $\tilde B_2$
appearing in (\ref{1e}). However, there is an ambiguity in this definition. In
general, we must allow
\be
\frac{1}{\tilde \ap} \tilde B_2({\rm supergravity}) =
\frac{1}{\tilde \ap} \tilde B_2({\rm dual ~~ string}) +
\frac{m}{\ap} B_2({\rm string})
\la{111}
\ee
where, in order to preserve the quantization condition on the WZW term, $m$
must be some integer. We find from (\ref{35xy}) that the correct result is
obtained by the choice\footnote{As we shall see
in Appendix B this admits the $D = 10$ interpretation of $-p_1/{24}$ where
$p_1$
is the Pontryagin number of $K3$ given in (\ref{23}).} $m = 2$. In this case,
\bea
d \tilde H_3({\rm supergravity}) &=& \tilde \ap (2\p)^2\, \tilde X_4 \nn \\
&=& \frac{\tilde \ap}{4}
\Bl [ -2\tr F_{SO(28)}{}^2 +44\tr F_{SU(2)}{}^2 - \tr R^2 \Br ]
\la{11yz}
\eea
This modification to the dual string Bianchi identity is thus a classical dual
string effect (tree level in the $D = 6$ dual string expansion). This is
confirmed by noting that there is no dilaton dependence and that $d \tilde H_3$
is independent of $\kappa^2$ and that the $2\pi$ factors cancel out. Once
again, by supersymmetry,
the same combination of Yang-Mills field strengths and gravitational curvatures
appearing in
(\ref{11yz}) also appears in the dual $D=6$ action
\bea
\tilde S^{(0)}{}_{YM}=\frac{1}{2\kappa^2}\int d^6 x\sqrt{-\tilde
g}e^{2\Phi}\frac{\tilde
\ap}{8}\tilde t^{\mu\nu\rho\sigma} \Br[ -2\tr
F_{SO(28)}{}_{\mu\nu}F_{SO(28)}{}_{\rho\sigma}\nn\\
 + 44\tr
F_{SU(2)}{}_{\mu\nu}F_{SU(2)}{}_{\rho\sigma}
 - \tr R_{\mu\nu} R_{\rho\sigma} +...\Br ]
\la{11yza}
\eea
where $\tilde t^{\mu\nu\rho\sigma}$ is obtained from $t^{\mu\nu\rho\sigma}$ of
(\ref{t}) by
replacing $g_{\mu\nu}$ with $\tilde g_{\mu\nu}$. Now here
is the crucial step: using the duality conditions (\ref{metric}), (\ref{e}),
(\ref{3}) and the
Dirac quantization rule (\ref{6a}) with $n = 1$, this classical modification to
dual
string Bianchi identity (\ref{11yz}) and tree level action
(\ref{11yza}) are seen to be identical to the
string one-loop correction to the fundamental string field equations (\ref{17})
and action
(\ref{17b}), respectively.  This is the main result of the paper.

\section{Conclusion}

According to string/string duality, the Green-Schwarz factorization of the
$D=6$ spacetime anomaly
polynomial $I_8$ into $X_4\tilde X_4$ means that just as $X_4$ is the
$\sigma$-model anomaly
polynomial of the fundamental string worldsheet so $\tilde X_4$ should be the
corresponding
polynomial of the dual string worldsheet. To test this idea we have performed a
classical dual string calculation of $\tilde{X}_4$ and found agreement
with the quantum fundamental string result. Moreover, as we discuss in the
Appendices, the same
result for $\tilde X_4$ can be obtained by starting with the $D=10$ fivebrane
and using
\be
\tilde X_4=\int_{K3} \tilde X_8
\ee
where $\tilde X_8$ is the anomaly polynomial of the $d=6$ fivebrane worldvolume
calculated in a previous paper \cite{dixon} on string/fivebrane duality. This
therefore provides an {\it a posteriori} justification for assumptions made in
\cite{dixon} on the {\it gravitational} contribution to $\tilde X_8$, although
we agree with \cite{izquierdo,blum} that an {\it a priori} justification is
still lacking.  However, we disagree with these authors concerning their
criticism of \cite{dixon} that the gauge fermions of the covariant heterotic
fivebrane cannot belong to the $32$ of $SO(32)$.  They claim that this is
inconsistent with Strominger's result \cite{strominger} that the fermions of
the gauge-fixed solitonic fivebrane belong to the $(2,28)$ of $SU(2) \times
SO(28)$.  The fivebrane soliton, with its $SU(2)$ instanton in the four
transverse directions, breaks the spacetime plus internal symmetry $SO(1,9)
\times SO(32)$ to $SO(1,5) \times SU(2) \times SO(28)$. (Note that the
embedding $SO(1,9) \times SO(32) \supset SO(1,5) \times SO(29)$
\cite{izquierdo} is not the minimal embedding and corresponds to two minimal
instantons \cite{blum}). Assuming that the fermions in the unbroken phase
transform as a $(16,32)$ is entirely consistent with a $(4,2,28)$ in the broken
phase.

Despite the success of our consistency check, however, some questions remain
unresolved. Firstly, we have concentrated on the $D=6$ string obtained from the
$SO(32)$ heterotic string in $D=10$.  It would be interesting to repeat the
exercise for the $E_8 \times E_8$ heterotic string and for the Type $IIB$
string but we have not yet done so. Secondly, the dual string seems very
different from the fundamental string. In particular, whereas for the
fundamental string the number of left and right moving gauge fermions
$(n_L,n_R)$ is $(20,4)$ and the Kac-Moody levels are $k_{SO(28)}=1$ and
$k_{SU(2)}=1$, for the dual string we have $(60,44)$ and $\tilde k_{SO(28)}=4$
and
$\tilde k_{SU(2)}=-20$. Although this dual heterotic string still has
$n_L-n_R=16$,
there seems to be a problem with conformal invariance since we get the wrong
central charge.  There are some caveats to be made in this connection, however.
 For the purpose of calculating $\sigma$-model anomalies,  we have focussed
only on the massless states of the dual string worldsheet.  There will also be
massive Kaluza-Klein modes on the worldsheet coming from compactifying the
fivebrane worldvolume on $K3$ and these could contribute to the conformal
anomaly.  Nor should we forget that the fourbrane, threebrane and membrane are
also present, even though they make no contribution to the $\sigma$-model
anomaly. Thirdly,  the $SU(2)$ Yang-Mills kinetic energy terms for the dual
supergravity action (\ref{11yza}) also appears to have the wrong sign, related
to the wrong sign
for the corresponding Kac-Moody level.  However, it should be borne in mind
that the fundamental
action and the dual action are equivalent up to duality transformations and
that $S^{(0)}{}_{YM}$
and $\tilde S^{(0)}{}_{YM}$ contribute to both.  The difference lies only in
the
loop expansion: in particular what is tree-level for the fundamental string
Yang-Mills kinetic energy is one-loop for the dual string ($\tilde
S^{(0)}{}_{YM}=S^{(1)}{}_{YM}$), and vice-versa. This may be verified by
converting to the appropriate $\sigma$-model metric and counting powers of
$e^{\Phi}$. On the subject of loop expansions, it was conjectured in \cite{lu2}
that the number of fivebrane loops $\tilde L_6$ was related to the Euler number
of the fivebrane worldvolume $\tilde \chi_{6}$ by the formula $\tilde
\chi_{6}=2(1-\tilde L_6)$ in just the same way that the number of string loops
$L_2$ is related to the Euler number of the string worldsheet $\chi_2$ by the
formula $\chi_2=2(1-L_2)$.  In the present context $\tilde M_6=\tilde M_2
\times K3$ and the Euler number is given by
$\tilde \chi_6=48(1-\tilde L_2)$ and hence
\be
\tilde L_6=24\tilde L_2-23
\ee
In summary, the success in achieving the right anomaly has still left
unanswered the question of
whether the dual theory (string plus membrane, threebrane and fourbrane) is
itself quantum
mechanically consistent.

Perhaps the most important lesson to be learned is that the nature of the dual
string depends crucially on the compactification.  The inconsistencies (if
there are any) may thus be a blessing in disguise: perhaps the requirement that
both the fundamental {\it and} the dual string be consistent will thus provide
a non-perturbative way of narrowing down the range of allowed superstring
vacua.  One might even entertain the idea that
in the perfect vacuum the dual string is identical to
the fundamental string.  For example, requiring that the dual string be chiral
with non-vanishing $\tilde X_4$ means that $I_8$ must be non-vanishing and
hence that the spacetime theory must be chiral.  This would rule out toroidal
compactification, for example, which is not ruled out
perturbatively\footnote{There is something rather asymmetrical about toriodal
compactification to four spacetime dimensions: the target space duality of the
fundamental string is $O(6,22;Z)$ and that of the dual string $SL(6,Z) \times
SL(2,Z)$ \cite{lu1,font,schwarz4,binetruy,khuri2}.}. It is obviously of
interest in
this context to see whether four spacetime dimensions is superior to six.

   \section{Acknowledgements}

We are grateful to John Dixon and Ergin Sezgin for many useful discussions.
Conversations with
Jeffrey Harvey, Jim Liu and Paul Townsend are also gratefully acknowledged.

\appendix

\section{Review of string/fivebrane test in $D=10$}

$D=10$ supergravity-Yang-Mills admits two anomaly-free \cite{green,salam}
formulations: one with a three-form field strength $H_3$
\cite{bergshoeff,chapline} and the other with a seven-form field strength
$\tilde H^{'}{}_7$ \cite{chamseddine}. They are related by Poincare duality:
\be
\tilde H^{'}{}_7 = e^{-\hat\Phi}\, {\ast H_3}
\la{2}
\ee
where $\hat\Phi$ is the $D=10$ dilaton and $\ast$ denotes the Hodge dual using
the
canonical metric $g^c{}_{MN}$ $(M=0,1,...,9)$. Taking the exterior derivative
of both sides of
(\ref{2}) reveals that the roles of field equations and Bianchi identities of
the
three-form version of supergravity are interchanged in going to the seven-form
version. The former corresponds to the field theory limit of the heterotic
string while the latter is conjectured to be the field theory limit of an
extended object dual to the string: the ``heterotic fivebrane''
\cite{duff1,strominger}. Just as the 2-form potential $B_{MN}$
couples to the $d=2$ string worldsheet via the term
\be
S_2 = {1 \over 2\p \ap} \int_{M_2} d^2 \x \, {1\over 2} \e^{ij} \pa_i x^M \pa_j
x^N B_{MN}
 = {1 \over 2\p \ap} \int_{M_2} B_2
\la{3a}
\ee
where $\x ^i$ ($i=1,2$) are the worldsheet coordinates and $(2\p\ap)^{-1}$ is
the string tension, so the 6-form potential $\tilde B_{MNPQRS}$
($M=0,1,\ldots,9$) couples to the $\tilde d=6$ fivebrane worldvolume via the
term
\bea
\tilde S_6 &=& {1 \over (2\p)^3 \bp} \int_{\tilde M_6} d^6 \tilde \x \, {1\over
6!} \e^{ijklmn} \pa_i x^M \pa_j x^N \pa_k x^P \pa_l x^Q \pa_m x^R \pa_n x^S
\tilde B_{MNPQRS} \nonumber \\
 &=& {1 \over (2\p)^3 \bp} \int_{\tilde M_6} \tilde B_6
\la{4}
\eea
where $\tilde \x^i$ ($i=1,\ldots,6$) are the worldvolume coordinates and
$[(2\p)^3\bp]^{-1}$ is the fivebrane tension. The two tensions obey the
Dirac quantization rule \cite{lu2}
\be
2\k_{10}^2 = n(2\p)^5 \ap\bp,~~~~~~~~~~~~~~~n=integer
\la{5}
\ee
where $\k_{10}^2$ is the $D=10$ gravitational constant. To within $O(\ap)$
Chern--Simons corrections to be
discussed below, $H_3$ is the curl of $B_2$
\be
H_3=dB_2 + O(\ap)
\la{6x}
\ee
Similarly, up to $O(\bp)$ Chern--Simons corrections to be discussed below,
$\tilde H{}_7$ is the curl of $\tilde B_6$.
\be
\tilde H^{}_7 = d\tilde B^{}_6 + O(\bp)
\la{7}
\ee
We are tempted to identify $\tilde H{'}{}_7$ appearing in (\ref{2}) with
$\tilde H{}_7$ appearing in (\ref{7}). However, as pointed out in \cite{dixon},
this identification needs to be modified when we include the gravitational
Chern--Simons corrections which are of higher order in the low-energy
expansion than those of Yang--Mills. Accordingly, we wrote \cite{dixon}
\be
\tilde H^{'}{}_7 =\tilde H_7 - \frac{1}{48}\, \frac{\bp}{\ap}\, \tr R^2\, H_3
\la{8}
\ee
Here $R = d\omega + \omega^2$ and the Lorentz connections $\omega = \omega_M\,
dx^M$ are in the vector representation. The choice of coefficient $1/48$ is
significant and we shall return to this later on.

In \cite{dixon}, the $O(\ap)$ corrections to $H_3$ and
the $O(\bp)$ corrections to $\tilde H^{'}{}_7$ were examined as a test of the
string/fivebrane
duality conjecture.
One begins with the observation of that duality mixes up
string and fivebrane loops: what is a one loop effect for the string might be
a tree level effect for the fivebrane, and vice versa. It is this loop mixing
which allows us to test
string/fivebrane duality, in spite of our ignorance of how to quantize the
fivebrane. If duality is correct, we should be able to reproduce string loop
effects from tree-level fivebranes!
\par
To see this, let us first consider the well-known $SO(32)$ Yang--Mills and
Lorentz
Chern--Simons corrections to $H_3$. Let us define $F = dA + A^2$
where the gauge fields
$A = A_M\, dx^M$ are matrices in the fundamental representation
of $SO(32)$. Let us further define
\bea
I_4 &=& \frac{1}{(2\p)^2} \Bl [ -\frac{1}{2}\,\tr F^2
+ \frac{1}{2}\,\tr R^2 \Br ] \nonumber \\
d \w_{3} &=& I_{4} \nonumber \\
 \d \w_{3} &=& d \w^1_{2}
\la{9b}
\eea
then the action $S_2$ can be modified so as to be both gauge invariant and
Lorentz provided \cite{hull}
\be
\d B_2 = \frac{1}{2} \ap (2\p)^2 \w^1_2
\la{10a}
\ee
and hence the gauge invariant field strength is given by
\bea
H_3 &=& dB_2 - \frac{1}{2} \ap (2\p)^2 \w_3 \nonumber \\
dH_3 &=& - \frac{1}{2} \ap (2\p)^2 I_4
\la{11x}
\eea
This modification to the Bianchi identity is thus a classical string
effect (i.e. tree level in the $D=10$ string loop expansion).
This is confirmed by the observation that there is no dilaton dependence and
that $dH_3$ is independent of $\k_{10}^2$ and that the $2\pi$ factors cancel
out.

Next we turn to the Green--Schwarz anomaly cancellation mechanism \cite{green}.
The anomaly receives contributions
from the gravitino and the dimension 496 adjoint representation gauginos of the
$D=10$ theory, both of which are Majorana--Weyl. As emphasized by Green and
Schwarz \cite{green}, the
miracle of $SO(32)$ is that $I_{12}$ factorizes:
\be
\frac{1}{2} I_{12} = X_4 \, \tilde X_8
\la{12}
\ee
where
\bea
X_4=\frac{1}{2} I_{4}
\la{13}
\eea
and
\bea
\tilde X_8 &=& \frac{1}{(2\p)^4}\,
 \Bl [ \frac{1}{24} \,\tr F^4 - \frac{1}{192}\,
\tr F^2 \, \tr R^2 + \frac{1}{768}\,(\tr R^2)^2 + \frac{1}{192}\, \tr R^4
\Br ]
\la{14x}
\eea
The factors of $1/2$ in front of $I_{12}$ and $I_{4}$ arise because
both $D=10$ spacetime fermions and the $d=2$ worldsheet fermions are Majorana.
Defining $\tilde \omega^{'}{}_7$ by $\tilde X_8=d \tilde \omega^{'}{}_7$, the
consistent anomaly is then cancelled by adding to the effective action
\be
\D \G_2 = - 2\p \int_{M_{10}} \Bl ( \frac{1}{ \ap (2\p)^2} \, B_2\,
 \tilde X_8 + \frac{1}{3}\, \w_3 \,\tilde \w^{'}{}_7 \Br )
\la{15x}
\ee
and recalling the transformation rule for $B_2$ given in (\re{11x}). Now for
$B_2$ normalized as in (\re{a}), its kinetic term is
\be
\G_2 = - \frac{1}{2\k_{10}^2} \int_{M_{10}} \frac{1}{2} e^{-\f}\, H_3 \wedge
\ast H_3
\la{16}
\ee
and hence the addition of (\re{15x}) modifies the field equation to
\be
d(e^{-\f}\, \ast H_3) = \frac{2\k_{10}^2}{\ap (2\p)}\, \tilde X_8
\la{17a}
\ee
This modification to the field equations is thus a string one-loop effect.
This is confirmed by the dilaton dependence on the left hand side and by noting
that the right hand side is linear in $\k_{10}{}^2$
and involves a factor $1/({2\pi})^5$ appropriate to a
one-loop Feynman integral in $D=10$.
\par
So far, all our considerations started with the string worldsheet. The acid
test for string/fivebrane duality is to reproduce (\re{17a}) starting from the
fivebrane worldvolume. Let us define
\bea
\tilde I_8 &=&\frac{1}{(2\p)^4}\, \Bl [ \frac{1}{24}\,\tr F^4 -
\frac{1}{96}\,\tr F^2 \, \tr R^2 + \frac{5}{768}\,(\tr R^2)^2 +
\frac{1}{192}\,\tr R^4 \Br ]\nonumber \\
d \tilde \w_{7} &=& \tilde I_{8} \nonumber \\
 \d \tilde \w_{7} &=& d \tilde \w^1_{6}
\la{18}
\eea
Then, as shown in \cite{dixon} the action $\tilde S_6$ of section 2 can be
modified so as to be both gauge invariant and Lorentz invariant, provided
\be
\d\tilde B_6 = - \bp (2\p)^4\tilde\w^1_6
\la{19}
\ee
and hence the gauge invariant field strength is given by
\bea
\tilde H_7 &=& d\tilde B_6 + \bp (2\p)^4 \tilde \w_7 \la{30a} \\
d\tilde H_7 &=& \bp (2\p)^4 \tilde I_8
\la{20}
\eea
Note that $\tilde I_8$ is not quite identical to $\tilde X_8$ but
\bea
\tilde I_8 -\tilde X_8=\frac{1}{48}\frac{1}{(2\pi)^2}trR^2 X_4
\la{21}
\eea
and so invoking (\ref{8}), (\ref{11x}), (\ref{12}) and (\ref{21}), the
gauge-invariant field strength $\tilde H'{}_7$ satisfies
\be
d \tilde H'{}_7= \bp (2\p)^4\,\tilde X_8
\la{22}
\ee
This modification to the Bianchi identity is thus a classical fivebrane effect
(tree-level in the $D=10$ fivebrane loop expansion). This is confirmed by
noting that there is no dilaton dependence and that $d \tilde H'{}_7$
it is independent of $\k_{10}{}^2$ and that the factors of $2\pi$ cancel.
Now here is the crucial step: using the duality condition (\ref{2}) and the
Dirac quantization rule (\ref{5}), this classical modification to the fivebrane
Bianchi identity (\ref{22}) is seen to be identical to the string one-loop
correction to the string field equations (\ref{17a}).

\section{$D=6$ results from $D=10$}

In this Appendix, we reproduce the $D=6$ dual string results of section 5
starting from the $D=10$ dual fivebrane results of \cite{dixon} reviewed
in Appendix A.

First we note that the dual string worldsheet $\tilde M_{2}$ is obtained by
compactification on $K3$ of the dual fivebrane worldvolume
$\tilde M_{6} = \tilde M_{2} \times K3$. In particular, the 2-form of
$\tilde B_2$ of (\ref{1e}) is related to the 6-form $\tilde B_6$ of (\ref{4})
by
\be
{1 \over 2\p \tilde \ap}{\tilde B_2} ={1 \over (2\p)^3 \bp} \int_{K3} \tilde
B_6
\la{b1}
\ee
where
\be
{1 \over 2\p \tilde \ap} = {1 \over (2\p)^3 \bp} V
\la{b2}
\ee
and $V$ is the volume of $K3$. Similarly,
\be
\frac{1}{\k_{10}^2} = \frac{V}{\k^2}
\la{b3}
\ee
So the $D=6$ Dirac quantization rule (\ref{6a}) follows from the $D=10$ rule
(\ref{5}). As a consequence of (\ref{b1}), we have
\be
d {\tilde H_3} = {\tilde \ap \over (2\p)^2 \bp} \int_{K3} d {\tilde H_7}
={\tilde \ap} (2\p)^2 \int_{K3} {\tilde I_8}
\la{b4}
\ee
on using (\ref{22}). This provides us with the consistency check. Under the
compactification of section 3,
\bea
\tr F_{SO(32)}{}^2 &=& \tr F_{SO(28)}{}^2 + 2\tr F_{SU(2)}{}^2 + 2\tr F_{0}{}^2
\nn \\
\tr F_{SO(32)}{}^4 &=& \tr F_{SO(28)}{}^4 + 2\tr F_{SU(2)}{}^4
+ 6\tr F_{SU(2)}{}^2 \tr F_{0}{}^2 + 2\tr F_{0}{}^4
\nn \\
\tr R_{SO(1,9)}{}^2 &=& \tr R_{SO(1,5)}{}^2 + \tr R_{0}{}^2 \nn \\
\tr R_{SO(1,9)}{}^4 &=& \tr R_{SO(1,5)}{}^4 + \tr R_{0}{}^4
\la{b5}
\eea
Using the fact that $\tr R_{0}{}^2 = 2\tr F_{0}{}^2$, together with (\ref{23})
we can now perform the integration bearing in mind that only $F_0$ and $R_0$
lie in the $K3$ subspace.
Thus a 12th order polynomial with arbitrary coefficients $a, b, c, d, e$ yields
\bea
\frac{1}{(2\pi)^4} \, \frac{1}{24} \,\int_{K3} \Bl [ a\,\tr F^4 +
b\,(\tr F^2)^2 + c\,\tr F^2 \, \tr R^2 + d\,(\tr R^2)^2 +
e\,\tr R^4 \Br ] \nn \\
= \frac{1}{(2\pi)^2} \, \Bl [ 4(2b+c)\,\tr F_{SO(28)}{}^2
+ 4(3a + 4b +2c)\,\tr F_{SU(2)}{}^2 + 4(c+2d)\,\tr R_{SO(1,5)}{}^2
\Br ]
\la{b6}
\eea
In particular, from (\ref{14x}) and (\ref{18})
\bea
 \int_{K3} {\tilde I_8} &=& \frac{1}{2} {\tilde I_4} \nn \\
 \int_{K3} {\tilde X_8} &=& {\tilde X_4}
\la{b7}
\eea
in complete agreement with (\ref{11y}) and (\ref{11yz}).

It is now of interest to see how the $D=6$ shift (\ref{111}) follows from the
$D=10$ shift (\ref{8}). In particular, the factor ${1/48}$ was rather
mysterious
from $D=10$ point of view \cite{dixon}. Now we see the significance of $48$ in
the $D=6$ context as the Pontryagin number of $K3$, and the integer $m$ is
given
by
\be
m =-\frac{1}{48 (2\pi)^2}\int_{K3}tr R_{0}{}^2= - \frac{p_1}{24} = 2
\la{b8}
\ee
The above results (\ref{b6}), (\ref{b7}) and (\ref{b8}) thus provide an
{\it a posteriori}
justification for for the choice in \cite{dixon} of polynomial $\tilde I_8$ in
(\ref{18}) and shift in (\ref{21}). The point being that the coefficients
$c, d, e$ chosen in \cite{dixon} on the basis of an educated guess of the mixed
and gravitational fivebrane sigma-model anomalies, yield the well-known
gravitational sigma-model anomaly polynomial for the string as well as the
correct Yang--Mills terms.

In a similar way we may derive the terms in the $D=6$ dual action (\ref{11yza})
quadratic in the
Yang-Mills field strengths and gravitational curvature.  We begin with the
tree-level heterotic
fivebrane action which, as discussed in \cite{lu4}, must be {\it quartic} in
the field strengths:
\bea
\tilde S^{(0)}{}_{YM}=\frac{1}{2\kappa^2}\int d^{10} x\sqrt{-\tilde
g}e^{{2\hat\Phi/3}}\frac{ \bp}{24}\tilde t^{IJKLMNPQ} \Br[ \tr
F_{IJ}F_{KL}F_{MN}F_{PQ}\nn\\ -\frac{1}{8}trF_{IJ}F_{KL}trR_{MN}R_{PQ}
+\frac{1}{32}trR_{IJ}R_{KL}trR_{MN}R_{PQ}
+\frac{1}{8}trR_{IJ}R_{KL}R_{MN}R_{PQ} +...\Br]
\la{b9}
\eea
where $\tilde g_{MN}=e^{-\Phi/6}g^c{}_{MN}=e^{-2\Phi/3}g^{}_{MN}$ is the
fivebrane $\sigma$-model
metric \cite{lu2} and where
\bea
t^{IJKLMNPQ}=-\frac{1}{2}(g^{IK}g^{JL}-g^{IL}g^{JK})(g^{MP}g^{NQ}-g^{MQ}g^{NP})
\nn\\
-\frac{1}{2}(g^{KM}g^{LN}-g^{KN}g^{LM})(g^{PI}g^{QJ}-g^{PJ}g^{QI})\nn\\
     -\frac{1}{2}(g^{IM}g^{JN}-g^{IN}g^{JM})(g^{KP}g^{LQ}-g^{KQ}g^{LP})\nn\\
+\frac{1}{2}(g^{JK}g^{LM}g^{PN}g^{QI}+g^{JM}g^{NK}g^{LP}g^{QI}+
g^{JM}g^{NP}g^{KQ}g^{LI}
+ permutations) \eea
Integrating over $K3$ as in (\ref{b6}) above yields the $D=6$ dual string
action
(\ref{11yza}).

Finally, we note that the fundamental string tree-level but $4$ worldsheet
loop contribution to
the effective action in $D=10$ takes the form \cite{pope}

\bea
\frac{1}{2\kappa^2}\int d^{10} x\sqrt{-
g}e^{-2\Phi}{ \ap}t^{IJKLMNPQ} \Br[ btr
F_{IJ}F_{KL}tr F_{MN}F_{PQ}\nn\\ -2btrF_{IJ}F_{KL}trR_{MN}R_{PQ}
+btrR_{IJ}R_{KL}trR_{MN}R_{PQ}
+etrR_{IJ}R_{KL}R_{MN}R_{PQ} \Br]
\eea
and hence integrating over $K3$ as in (\ref{b6}) above yields a vanishing
contribution to the $D=6$
string action (\ref{9c}) and so does not interfere with any of our previous
results.

\end{document}